\documentstyle[preprint,aps]{revtex}
\begin{document}
\draft
\title{Models for Chronology Selection}
\author{M. J. Cassidy and S. W. Hawking}
\address{Department of Applied Mathematics and Theoretical Physics,\\ University of
Cambridge, Silver Street, Cambridge, CB3 9EW, England}
\date{\today}
\maketitle
\begin{abstract}
In this paper, we derive an expression for the grand canonical partition function for
a fluid of hot, rotating massless scalar field particles in the Einstein universe. We 
consider the number of states with a given energy as one increases the angular 
momentum so that the fluid rotates with an increasing angular velocity. We find that
at the critical value when the velocity of the particles furthest from the
origin reaches the speed of light, the number of states tends to zero. We illustrate
how one can also interpret this partition function as the effective action for a 
boosted scalar field configuration in the product of three dimensional de Sitter space 
and $S^1$. In this case, we consider the number of states with a fixed linear momentum
around the $S^1$ as the particles are given more and more boost momentum. At the
critical point when the spacetime is about to develop closed timelike curves, 
the number of states again tends to zero. Thus it seems that quantum mechanics
naturally enforces the chronology protection conjecture by superselecting the
causality violating field configurations from the quantum mechanical phase space.  
\end{abstract}

\pacs{PACS Number(s):}

\narrowtext     

\section{Introduction}
\label{cvone}

It is generally believed that any attempt to introduce closed timelike curves (CTCs) 
into
the universe will fall foul of the chronology protection conjecture \cite{cpc}, which
states that the laws of Physics somehow conspire to prevent one from manufacturing time
machines. Early calculations supporting this conjecture concentrated on the behaviour
of the renormalised energy--momentum tensor $\langle T_{\mu\nu}\rangle$, which was
shown to diverge at the Cauchy horizon in a number of causality violating spacetimes.
A possible mechanism for enforcement of chronology
protection was therefore proposed as the back reaction of this divergent 
energy--momentum
on the spacetime geometry via the semi--classical Einstein equations. Of course, it
was hoped that the back reaction would be sufficiently strong enough to prevent the 
formation of CTCs. However, Kim and Thorne speculated \cite{kth} that if a full 
quantum theory of gravity were available, then one might find that the divergences
cut off at some appropriate invariant distance from the Cauchy horizon, thus
allowing the CTCs to form. Further doubts were cast when $\langle T_{\mu\nu}\rangle$
was calculated for scalar fields in two spacetimes with noncompactly generated Cauchy
horizons. Boulware \cite{boul} and Tanaka and Hiscock \cite{tanhis} both found that
for sufficiently massive fields in Gott space and Grant space respectively, 
$\langle T_{\mu\nu}\rangle$ could remain regular at the Cauchy horizon. More recently,
it has been shown that Hadamard states exist in Misner space (in 2 and 4 dimensions)
for which $\langle T_{\mu\nu}\rangle$ vanishes everywhere \cite{sush,mjc,kras}. Misner
space has a compactly generated Cauchy horizon and is therefore subject to the strong
theorems recently proved by Kay, Radzikowski and Wald \cite{krw}. Cramer and Kay
\cite{ckay} have applied these theorems to the Misner space example and showed that 
even if
there was no divergence as the Cauchy horizon was approached, $\langle T_{\mu\nu}
\rangle$ must necessarily be ill defined on the Cauchy horizon itself. They also argue
for similar behaviour in the noncompactly generated cases, but the fact that
the energy--momentum tensor fails to diverge shows that back reaction does not
enforce chronology protection. 

In this paper, we adopt a slightly different approach by focusing on the effective
action of a massless scalar field in a number of acausal spacetimes. From a 
formal point of view, the effective action is the fundamental
field--theoretic quantity, from which the energy--momentum
tensor is derived as a functional derivative with respect to the metric, so one would
hope that an analysis of this quantity would provide new insights into issues
of chronology protection. The effective action plays an important role in the 
Euclidean approach to quantum field theory on acausal
spacetimes \cite{swh1}. This approach can be used if some Euclidean space has an
appropriate Lorentzian causality violating analytic continuation. CTCs do not exist in
Euclidean space, so one can define a field theory on the Euclidean section, and then
analytically continue to obtain the results valid for the acausal
spacetime. In this formalism, one defines path integrals of the form
\begin{equation}
\label{pathint}
Z=\int{\cal D}[\phi] \exp\Bigl(-S[\phi]\Bigr)
\end{equation}
over field configurations $\phi$. Our aim is to provide thermodynamic and 
quantum cosmological interpretations to expressions of this type in the presence of 
causality violations. The main obstacle to any such interpretation comes from the fact 
that in general,
the effective action density $\ln Z(x)$ diverges to infinity at the polarised 
hypersurfaces of 
acausal spacetimes \cite{mjc}. Thus, if one was to construct a no--boundary amplitude
for some causality violating geometry, then it would appear that creation of the
universe was overwhelmingly favoured, contrary to ones intuitive hope for a strong
suppression. However, we will argue that the effective action in itself does not yield
the correct amplitude for creation, and that the true amplitude does indeed show
suppression of acausal geometries.

In section\ \ref{cvtwo}, we introduce two multiply connected Euclidean spaces, from
which one can obtain a variety of acausal Lorentzian spacetimes. When one tries to do 
physics in a typical multiply connected spacetime, it is
generally easier to work in the simply connected universal covering space, where
points are identified under the action of some discrete group of
isometries. The first example considered, therefore, is flat Euclidean space
with points identified under a combined rotation plus translation. By analytically
continuing the rotation to complex values ($\alpha\rightarrow a=i\alpha$), one can
obtain Grant's generalisation of Misner space \cite{grant}, which is just flat
Minkowski space with points identified under a combined boost in the $(x,t)$ plane and
orthogonal translation in the $y$ direction. This spacetime contains CTCs in the 
left and right wedges (because of the boost identification) and
is the covering space of the Gott spacetime \cite{gott}, which describes two
infinitely long  cosmic strings moving past each other at high velocity. One can also
find other acausal analytic continuations from this Euclidean section. We illustrate
how one obtains the `spinning cone' spacetime \cite{spin} and the metric for hot flat
space, rotating rigidly with angular velocity $\Omega$. In this latter spacetime, the
velocity of a co--rotating observer increases as one moves radially outward from the
axis of rotation and there will be acausal effects beyond the critical radius where 
this velocity reaches the speed of light.

The second example that we introduce is a new model, given by the Euclidean metric on
$R\times S^3$, also identified under a combined rotation and translation. The
basic reason for introducing this model is to provide a compact Euclidean space which
could, in principle, contribute to a no--boundary path integral. It should
not be surprising that the analytic continuations of this model have a causal
structure qualitatively similar to the flat space examples. Indeed, by allowing the
radius of the sphere to tend to infinity, one regains the periodically identified flat
Euclidean space. The acausal spacetime analogous to Grant space, obtained by
analytically continuing the rotation to a boost, is the product of three--dimensional
de Sitter space and the real line ($3dS\times R$), periodically identified under a
combined boost and translation.

In section\ \ref{cvthree}, we introduce a massless scalar field into these two
Euclidean models and calculate the renormalised energy--momentum
tensor in each case. Not surprisingly, we find that all the
components of $\langle T_{\mu\nu}\rangle$ diverge at the
Cauchy horizon and polarised hypersurfaces in all of the acausal analytic
continuations.

Section\ \ref{cvfour} is devoted to a calculation of the contributions to the
effective action which diverge when one analytically continues the parameters of the
Euclidean Einstein universe. In \cite{mjc}, a divergent contribution was derived from
the de Witt--Schwinger asymptotic expansion of the heat kernel $H(x,x',\tau)$ about
$\tau=0$ and here, this contribution is rederived by integrating the energy--momentum
tensor. When one integrates $\langle T_{\mu\nu}\rangle$, however, one also finds other
divergent contributions that do not appear when one calculates $\ln Z(x)$ directly from
the heat kernel expansion. Ultimately, though, the dominant divergence is the same as
before -- the effective action density diverges to infinity at the polarised 
hypersurfaces of the acausal analytic continuations. 

In section\ \ref{cvfive}, we give a physical interpretation to the results of the
previous section. Firstly, we consider a fluid of hot, rotating scalar particles in
the Einstein universe. The grand canonical partition function for these particles is
given by $\ln Z(x)$, with the parameter $\alpha$ continued to complex values so that the
fluid is rotating with angular velocity $\Omega= i{\alpha\over\beta}$, where $\beta$
is the inverse temperature. Energy and angular momentum are conserved quantities in
the Einstein universe, and we derive expressions for the energy of the particles and
also the angular momentum that is required to make the fluid rotate with an angular
velocity $\Omega$. At a critical angular velocity $\Omega={1\over r}$, where $r$ is
the radius of the Einstein universe, these expressions diverge which means that one 
would have to inject an infinite amount of angular momentum into the system if one 
wanted the velocity of the particles to reach the speed of light. Ultimately, however,
one is interested in the behaviour of the number of states with a given energy as the
angular momentum is increased. In order to keep the energy fixed, one must decrease
the temperature as more and more angular momentum is put in so that the angular
velocity of the particles approaches its critical value. One finds that the entropy of
the scalar particles diverges to minus infinity as the velocity of the particles
approaches the speed of light (or $\Omega\rightarrow{1\over r}$). Since the entropy is
just the logarithm of the number of states, one can conclude that there are no quantum
states available for speed of light rotation.

These results can be interpreted analogously if one analytically continues the
Euclidean section to obtain $3dS\times S^1$, the product of three dimensional de
Sitter space and the $S^1$ with length $\beta$. The conserved quantities are now
linear and boost momentum so this time, one wants to consider the number of states
with a fixed linear momentum as one gives the particles more and more boost momentum.
There is a critical boost at which the spacetime will develop CTCs (when $a=\beta/r$),
but the amount of boost momentum that is needed to achieve this is once again infinite
and the entropy diverges to minus infinity at this critical value. In this case,
therefore, there are no quantum states available for these causality violating field
configurations.

The no--boundary amplitude $\Psi_m$ which describes the creation of the spacetime
$3dS\times S^1$ from nothing is also constructed. $\Psi_m^2$ is the microcanonical
partition function, or density of states, which tends to zero as one adjusts the
parameters so that the spacetime is about to develop CTCs. The amplitude $\ln \Psi_m$
is obtained from the original effective action by a Legendre transform, just as one
obtains the entropy from the partition function in a thermodynamic context. 
Thus the message of this paper is that it is possible to recover a sensible
interpretation of Euclidean path integrals in the presence of causality violation as
long as one focuses on the density of states. It seems highly likely that this
quantity will always tend to zero as one tries to introduce CTCs, thus enforcing the
chronology protection conjecture.   

\section{Periodically identified Euclidean spaces}
\label{cvtwo}

In this section, we illustrate how CTCs can be
introduced into a spacetime by identifying points under the action of a discrete group
of isometries. Consider the metric for flat Euclidean space in cylindrical polar
coordinates
\begin{equation}
\label{basic}
ds^2=d\tau'^2 +dr'^2 +r'^2d\phi'^2 + dz'^2
\end{equation}
where points are identified under a combined rotation and translation, {\it i.e.}
$(\tau', r', \phi', z')$ and $(\tau'+n\beta, r', \phi' + n\alpha,z')$ represent the 
same spacetime
point (where $n$ is some integer). The identification parameters appear explicitly in 
the metric when one makes the coordinate transformation
\begin{eqnarray}
\tau&=&\tau' - {\beta\phi'\over\alpha} \nonumber \\
r&=&r' \nonumber \\
\alpha\phi &=& \phi' \nonumber \\
z&=&z'
\end{eqnarray}
to obtain
\begin{equation}
\label{flat}
ds^2=(d\tau + \beta d\phi)^2 + dr^2 + \alpha^2 r^2 d\phi^2 + dz^2\space.
\end{equation}
The idea now is to analytically continue one of the parameters $\alpha$ or $\beta$ to
obtain the acausal Lorentzian spacetimes.
If we continue $\beta\rightarrow b=-i\beta$ and then set $t=-i\tau$, we obtain
\begin{equation}
ds^2=-(dt+bd\phi)^2 + dr^2 + \alpha^2 r^2 d\chi^2 + dz^2
\end{equation}
which is the `spinning cone' metric \cite{spin}, the spacetime produced by an 
infinitely long
string with angular momentum $b$. The condition for CTCs is $(-b^2
+ \alpha^2 r^2)<0$, so the causality violating region is just
$0<r<{b\over\alpha}$. The spinning cone metric is singular along the axis of the
string and will not concern us further. It is interesting, however, to note that this
spacetime and Grant space (obtained in the next paragraph) are just different analytic
continuations of the same Euclidean metric.

Returning to equation (\ref{flat}), if one now continues $\alpha\rightarrow 
a= i\alpha$, one obtains the metric
\begin{equation}
\label{gra}
ds^2=-a^2r^2d\phi^2 + dr^2 + (d\tau + \beta d\phi)^2 + dz^2\space.
\end{equation}
Analytically continuing in $\alpha$ means that points are now identified under a 
combined
boost plus translation, so this metric is just that of Grant's generalised Misner space.
The condition for CTCs is $(\beta^2-a^2r^2)<0$. In other words, the CTCs
inhabit the region where $r>{\beta\over a}$. 
It should be stressed that the surface defined by $r={\beta\over a}$ is not the Cauchy
horizon for Grant space. If one thinks of the $(t,x)$ section of Minkowski space as 
being divided up into the usual four wedges, then for Grant space the CTCs are
confined to the $r>{\beta\over a}$ region of the left and right wedges but the Cauchy
horizon is defined by $t=\pm x$. The Cauchy horizon is in fact the
$n\rightarrow\infty$ limiting surface of a
family of $n$th polarised hypersurfaces. Physically, the $n$th polarised hypersurface
is defined as the set of points which can be joined to themselves by a
(self--intersecting) null geodesic which loops around the space $n$ times. In Grant
space, these surfaces are defined by the equation
\begin{equation}
2r^2(1-\cosh(na))+n^2\beta^2=0\space.
\end{equation}
In the limit as $n\rightarrow 0$, one obtains $r={\beta\over a}$, so one could say that
this surface is the zeroth polarised hypersurface but in light of the above 
definition, its physical interpretation is unclear. Bearing this in mind however, we 
shall continue to refer to it as the zeroth polarised hypersurface. Ordinary Misner
space is obtained when the translation parameter $\beta$ is zero. Misner space is
basically the Euclidean cosmic string metric, with the angular defecit parameter
continued to complex values.

A more familiar example of a spacetime containing CTCs is obtained from the original
metric (\ref{basic}) by the coordinate transformation
\begin{eqnarray}
\beta \tau&=&\tau' \nonumber \\
r&=&r' \nonumber \\
\phi&=&\phi' -{\alpha\tau'\over\beta} \nonumber \\
z&=&z'.
\end{eqnarray}
If one analytically continues $\beta\rightarrow b=-i\beta$ 
 in this case, one obtains
the metric for hot flat space, rotating rigidly with angular velocity
$\Omega={\alpha\over b}={a\over\beta}$
\begin{equation}
\label{hfm}
ds^2=-b^2 d\tau^2 + dr^2 + r^2(d\phi + \alpha d\tau)^2 + dz^2\space.
\end{equation}
In this metric, the Killing vector $\partial/\partial \tau$ becomes spacelike
 beyond the critical radius where the
velocity of a co--rotating observer exceeds the speed of light.

In a later section, we will be concerned with the possible contributions from acausal 
metrics to no--boundary amplitudes. The flat space examples above could not contribute
because their Euclidean section is noncompact. However, one can readily construct a 
compact example with spherical spatial sections, given by the Euclidean metric on 
$R\times S^3$ (the Euclidean Einstein universe)
\begin{equation}
ds^2= d\tau^2 + r^2\Bigl(d\chi^2 + \sin^2\chi(d\theta^2 + \sin^2\theta d\phi^2)\Bigr)
\end{equation}
where the points $(\tau,\chi,\theta,\phi)$ and $(\tau+m\beta,
\chi,\theta,\phi+m\alpha)$ are identified. Clearly one can analytically continue the
metric parameters to obtain acausal spacetimes analogous to the flat space
examples considered above. The spacetime obtained by just analytically continuing
$\alpha\rightarrow a=i\alpha$ is the product of three--dimensional de Sitter space
and the real line, periodically identified under a combined boost and translation, and
in this case the polarised hypersurfaces are defined by the equation
\begin{equation}
\sin^2\chi\sin^2\theta={1-\cosh\left(m\beta\over r\right)\over 1-\cosh(ma)}\space.
\end{equation}
The polarised hypersurfaces all coincide at the critical value when $a=\beta/r$ and
$\sin\chi\sin\theta=1$, and CTCs appear in the spacetime if $a$ is increased further. 
By taking the radius $r$ of the sphere to infinity, one obtains the flat space example 
as a limiting case.

\section{Scalar field energy--momentum tensor}
\label{cvthree}

Now consider placing a massless scalar field on the two identified Euclidean spaces
described in the previous section. To find the energy--momentum for either of these
spaces, one just applies the standard second order differential operator to the
appropriate Euclidean Green function. Analytically continuing at the end of the
calculation will yield the results for the acausal Lorentzian spacetimes. We first
consider the flat space example.

The renormalised Euclidean Green function for a massless scalar field on identified 
flat space is written using the method of images as
\begin{equation}
D(x,x')= {1\over4\pi^2}\sum_{\scriptstyle n=-\infty\atop\scriptstyle n\ne0}
^\infty
{1\over \sigma_n(x,x')},
\end{equation}
where 
\begin{equation}
\sigma_n(x,x')=r^2+r'^2-2rr'\cos(\phi-\phi'-n\alpha) + (\tau-\tau'-n\beta)^2+
(z-z')^2.
\end{equation}
The energy--momentum tensor is obtained by differentiating $D$ according to 
\begin{equation}
\langle T_{\mu\nu}\rangle=\lim_{x'\rightarrow x} \left[ {2\over3}{D}_{;\nu'\mu}
-{1\over3}{D}_{;\nu\mu} -{1\over6}g_{\mu\nu} {D}^{;\sigma'}{}_\sigma 
\right]
\end{equation}
and the individual components are given by
\begin{equation}
\langle T_{\tau\tau}\rangle = {1\over4\pi^2}\sum_{\scriptstyle n=
-\infty\atop\scriptstyle n\ne0}^\infty
{2\Bigl(\cos(n\alpha) + 2\Bigr)\over3\sigma_n(x,x)^2} -
{4n^2\beta^2\Bigl(\cos(n\alpha)+5\Bigr)\over3\sigma_n(x,x)^3}
\end{equation}
\begin{equation}
\langle T_{rr}\rangle = {1\over4\pi^2}\sum_{\scriptstyle n=-\infty\atop\scriptstyle 
n\ne0}^\infty
{2\Bigl( \cos(n\alpha) + 2\Bigr)\over3\sigma_n(x,x)^2}
\end{equation}
\begin{equation}
\langle T_{\phi\phi} \rangle = {1\over4\pi^2}\sum_{\scriptstyle n=
-\infty\atop\scriptstyle 
n\ne0}^\infty {2r^2\Bigl( \cos(n\alpha) +
2\Bigr)\over3\sigma_n(x,x)^2}\left[ -3 + {4n^2\beta^2\over\sigma_n(x,x)}\right]
\end{equation}
\begin{equation}
\langle T_{zz}\rangle = {1\over4\pi^2}\sum_{\scriptstyle n=-\infty\atop\scriptstyle 
n\ne0}^\infty
{2\Bigl(\cos(n\alpha) +
2\Bigr)\over3\sigma_n(x,x)^2} -
{4n^2\beta^2\Bigl(\cos(n\alpha)-1\Bigr)\over3\sigma_n(x,x)^3}
\end{equation}
\begin{equation}
\langle T_{\tau\phi}\rangle  = {1\over4\pi^2}\sum_{\scriptstyle n=
-\infty\atop\scriptstyle n\ne0}^\infty
-{8n\beta r^2
\sin(n\alpha)\over\sigma_n(x,x)^3}
\end{equation}

These results are valid on the Euclidean section. One obtains the energy--momentum
components for generalised Misner space (reproducing Grant's results) by analytically
continuing $\alpha\rightarrow a=i\alpha$. We also note that continuing in $\beta$ (and
$\tau$)
yields the energy--momentum for the spinning cone.

 The appropriate Green function for a massless scalar field on $R\times S^3$
identified under a combined rotation and translation is
\begin{equation}
D(x,x')={1\over 4\pi^2r}\sum_{m=-\infty}^{\infty}\sum_{n=-\infty}^{\infty}{\left(
{s_m+2\pi nr}\right)\over \sin({s_m\over
r})}{1\over\lambda_{mn}(x,x')}
\end{equation}
where $\lambda_{m\pm n}(x,x')=(\tau-\tau'-m\beta)^2 + (s_m\pm 2\pi nr)^2$ and $s_m=r
\cos^{-1}(\cos\chi \cos\chi' + \sin\chi\sin\chi'(\cos\theta 
\cos\theta' + \sin\theta \sin\theta'\cos(\phi-\phi'-m\alpha)))$.
By combining terms of positive and negative $n$, one can write $D(x,x')$ as the series
\begin{equation}
{1\over4\pi^2r}\sum_{m=-\infty}^{\infty}\sum_{n=-\infty}^{\infty}
{s_m\over \sin\left(s_m\over r\right)}{f_{mn}\over\lambda_{mn}\lambda_{m-n}}
\end{equation}
where $f_{mn}=\left(\tau-\tau'-m\beta\right)^2 + \left(s_m +2\pi nr\right)
\left(s_m-2\pi nr\right)$. The Green function is renormalised by
dropping the $n=0, m=0$ term in the sum, as this term is the only divergent one as 
the points are brought together. It is also convenient to separate the Green function 
as $D=D_1 + D_2$, where $D_1$ is the $m=0,\sum_n$ part of the Green function. 
In the limit as $x'\rightarrow x$, $D_1$  is given
by \cite{dc}
\begin{equation}
\lim_{x'\rightarrow x}D_1 =- {1\over48\pi^2 r^2}.
\end{equation}
This is just the value for the Einstein universe without identifications. 
The remainder of the Green function can be written as
\begin{equation}
D_2 = {1\over4\pi^2 r}\sum_{\scriptstyle m=-\infty\atop\scriptstyle m\ne0}
^\infty
\sum_{n=-\infty}^\infty {s_m\over16\pi^4 r^4 \sin\left(s_m\over r\right)}
{f_{mn}\over
\left(n+z_1\right)\left(n-z_1\right)\left(n+ z_1^*\right)\left(n-z_1^*\right)}
\end{equation}
where the complex quantity 
\begin{equation}
z_1={s_m + i(\tau-\tau'-m\beta)\over2\pi r}\space.
\end{equation}
The sum over $n$ can be evaluated using the method of residues. We find that
\begin{eqnarray}
D_2 &=& {1\over16\pi^2 r^2}\sum_{\scriptstyle 
m=-\infty\atop\scriptstyle m\ne0}^\infty 
{\cot\left(\pi z_1\right) + \cot\left(\pi z_1^*\right)\over\sin\left(s_m\over r\right)}
 \nonumber \\
&=&{1\over16\pi^2 r^2}\sum_{\scriptstyle m=-\infty\atop\scriptstyle m\ne0}^\infty
{1\over \sin\left({s_m + i(\tau-\tau'-m\beta)\over2r}\right)\sin\left({s_m-
i(\tau-\tau'-m\beta)\over2r}\right)} \nonumber \\
&=&{1\over8\pi^2 r^2}\sum_{\scriptstyle m=-\infty\atop\scriptstyle m\ne0}^\infty
{1\over \cosh\left({\tau-\tau'-m\beta\over r}\right) - \cos\left(s_m\over r\right)}
\end{eqnarray}
which represents $D_2$ as a sum over ordinary
Einstein Green functions, as one might expect. Furthermore, this quantity has already 
been
renormalised, so all that remains is to apply the standard formula to calculate the
energy--momentum

\begin{equation}
\langle T_{\mu\nu}\rangle = \lim_{x'\rightarrow x}\biggl({2\over3}D_{;\nu'\mu} - 
{1\over3}D_{;\nu\mu} 
-{1\over6}g_{\mu\nu}
D^{;\sigma'}{}_\sigma
+{1\over3}g_{\mu\nu}D^{;\sigma'}{}_{\sigma'} +
{1\over6}(R_{\mu\nu}-{1\over2}Rg_{\mu\nu})D\biggr).
\end{equation}
 The individual components are

\begin{eqnarray}
\langle T_{\tau\tau}\rangle=-{1\over480\pi^2r^4}&+&{1\over24\pi^2 r^2}
\sum_{\scriptstyle m=-\infty\atop\scriptstyle m\ne0}^\infty
\left\{{\cosh\left(m\beta\over r\right) + \cos(m\alpha)+1\over r^2\sigma_m(x)^2}\right.
\nonumber \\
&{}&\left.+{2\Bigl(1+\cos(m\alpha)\Bigr)\left(1-\cosh\left(m\beta\over r\right)
\right) -
4\sinh^2\left(m\beta\over r\right)\over r^2\sigma_m(x)^3}\right\}
\end{eqnarray}
\begin{eqnarray}
\langle T_{\chi\chi}\rangle={1\over1440\pi^2r^2}&+&{1\over24\pi^2 r^2}
\sum_{\scriptstyle m=-\infty\atop\scriptstyle m\ne0}^\infty 
\left\{{\cosh\left(m\beta\over r\right) + \cos(m\alpha)+1\over \sigma_m(x)^2}\right.
\nonumber \\
&{}&\left.-{2\cos^2\theta\Bigl(1-\cos(m\alpha)\Bigr)\left(1-\cosh\left(m\beta\over
r\right)\right)\over\sigma_m(x)^3}\right\} 
\end{eqnarray}
\begin{equation}
\langle T_{\chi\theta}\rangle={1\over24\pi^2 r^2}\sum_{\scriptstyle 
m=-\infty\atop\scriptstyle m\ne0}^\infty
{2\sin\chi\cos\chi\sin\theta\cos\theta\Bigl(1-\cos(m\alpha)\Bigr)\left(1-\cosh\left(
m\beta\over r\right)\right)\over\sigma_m(x)^3}
\end{equation} 
\begin{equation}
\langle T_{\tau\phi}\rangle=-{1\over8\pi^2 r^2}\sum_{\scriptstyle 
m=-\infty\atop\scriptstyle m\ne0}^\infty
{2\sin^2\chi\sin^2\theta \sinh\left(m\beta\over r\right)\sin(m\alpha)\over r
\sigma_m(x)^3}
\end{equation}

\begin{eqnarray}
\langle
T_{\theta\theta}\rangle={\sin^2\chi\over1440\pi^2r^2}&+&{\sin^2\chi\over24\pi^2 r^2}
\sum_{\scriptstyle m=-\infty\atop\scriptstyle m\ne0}^\infty 
\left\{{3-\cosh\left(m\beta\over r\right) + \cos(m\alpha)\over\sigma_m(x)^2}\right.
\nonumber \\
&+&\left. {2\left(1-\cosh\left(m\beta\over r\right)\right)^2 -
2\sin^2\theta\Bigl(1-\cos(m\alpha)\Bigr)\left(1-\cosh\left(m\beta\over r\right)\right)
\over\sigma_m(x)^3}\right\}
\end{eqnarray}
\begin{eqnarray}
\langle T_{\phi\phi}\rangle
={\sin^2\chi\sin^2\theta\over1440\pi^2r^2}&+&{\sin^2\chi\sin^2\theta\over24\pi^2 r^2}
\sum_{\scriptstyle m=-\infty\atop\scriptstyle m\ne0}^\infty
\left\{{-\left(\cosh\left(m\beta\over r\right) + 3\cos(m\alpha) +
5\right)\over\sigma_m(x)^2}\right.
\nonumber \\
&+&\left.{2\sinh^2\left(m\beta\over r\right)
-4\Bigl(1+\cos(m\alpha)\Bigr)\left(1-\cosh\left(m\beta\over r\right)
\right)\over\sigma_m(x)^3}\right\}
\end{eqnarray}
where $\sigma_m(x)=\cosh\left(m\beta\over
r\right)-1+\sin^2\chi\sin^2\theta\Bigl(1-\cos(m\alpha)\Bigr)$. Clearly all of these 
components diverge at the $n$th polarised hypersurfaces if one
analytically continues $\alpha\rightarrow a=i\alpha$.

\section{Divergent Contributions to the Effective Action}
\label{cvfour}

As we stated in the introduction, the ultimate aim of this paper is to provide
sensible interpretations for path integrals of the form
\begin{equation}
Z=\int {\cal D}[\phi] \exp\Bigl(-S[\phi]\Bigr)
\end{equation}
in the presence of causality violations. The trouble is that if one calculates the
effective action density $\ln Z(x)$ for matter fields in an acausal spacetime, then 
the results of
\cite{mjc} suggest that $\ln Z(x)$ generally diverges to infinity at each of the $n$th
polarised hypersurfaces as one analytically continues the background so that the
Lorentzian section is about to develop CTCs. 

For example, consider the periodically identified Euclidean Einstein universe. From the
energy--momentum tensor calculated at the end of the previous section, one can define
the change in effective action induced by a metric perturbation $\delta g_{\mu\nu}$ as
\begin{equation}
\delta \ln Z={1\over2}\int g^{1\over2}\langle T^{\mu\nu}\rangle \delta g_{\mu\nu}
d^4 x\space.
\end{equation}
In this case, the metric perturbations arise by varying the parameter $\alpha$, so
that if one begins with the metric
\begin{equation}
ds^2=g_{\mu\nu}dx^\mu dx^\nu=d\tau^2 + r^2\left(d\chi^2 + \sin^2\chi\left(d\theta^2 +
\sin^2\theta\left(d\phi+ {\alpha\over\beta}d\tau\right)^2\right)\right)\space,
\end{equation}
then the perturbed metric $g_{\mu'\nu'}=g_{\mu\nu}+\delta g_{\mu\nu}$ is obtained from
the original one by the coordinate transformation
$\phi=\phi'+\tau'{d\alpha\over\beta}$. The only nonzero perturbations are $\delta
g_{\tau\phi}= r^2\sin^2\chi\sin^2\theta {d\alpha\over\beta}$ and $\delta
g_{\tau\tau}=2r^2\sin^2\chi\sin^2\theta{\alpha d\alpha\over\beta^2}$, which implies
that the total change in action is given by
\begin{equation}
\label{dellnz}
\delta\ln Z=\int g^{1\over2}\Bigl( \langle T_{\tau\phi}\rangle -
{\alpha\over\beta}\langle T_{\phi\phi}\rangle\Bigr){d\alpha\over\beta}d^4x\space.
\end{equation}

Integrating up with respect to $\alpha$ should yield the total effective action. The
first term contributes
\begin{equation}
\label{uvpart}
{1\over4\pi^2r^4}\sum_{m=1}^\infty {\left(m\beta\over
r\right)^{-1}\sinh\left(m\beta\over r\right)\over\sigma_m(x)^2}
\end{equation}
to the effective Lagrangian, which diverges at the polarised hypersurfaces when one
analytically continues $\alpha\rightarrow a=i\alpha$ to obtain $3dS\times R$, the
product of three--dimensional de Sitter space and the real line, periodically
identified under a combined boost and translation. One can take the
$r\rightarrow\infty$ limit to obtain the contribution to the effective Lagrangian in
Grant space (if one defines a new radial coordinate $r'=r\sin\chi\sin\theta$)
\begin{equation}
{1\over\pi^2}\sum_{m=1}^\infty {1\over
\left(2r'^2\Bigl(1-\cosh(ma)\Bigr)+m^2\beta^2\right)^2}.
\end{equation}
We note that the expressions obtained here agree with those of \cite{mjc}, obtained by
a different method. However, (\ref{dellnz}) indicates that there should be an
additional contribution to the effective Lagrangian, given by $I=-\int\langle
T_{\phi\phi}\rangle\alpha {d\alpha\over\beta^2}$. This can be integrated by parts to
obtain
\begin{equation}
I=-{1\over\beta^2}\left\{ {\alpha\int\langle T_{\phi\phi}\rangle d\alpha} -
{\int\left(\int\langle T_{\phi\phi}\rangle d\alpha\right)d\alpha}\right\}.
\end{equation}
The relevant integrals can be solved by successive application of the formulae
\cite{integrals}
\begin{eqnarray}
\int{A+B\cos x\over(a+b\cos x)^n}dx &=& {1\over(n-1)(a^2-b^2)}\int{(n-1)(Aa-Bb) -(n-2)
(Ab-Ba)\cos x\over(a+b\cos x)^{n-1}}dx \nonumber \\
&{}&- {(Ab-Ba)\sin x\over(n-1)(a^2-b^2)(a+b\cos x)^{n-1}}
\end{eqnarray}
and
\begin{equation}
\int{dx\over a+b\cos x}=
{2\over\sqrt{a^2-b^2}}\tan^{-1}\left\{(a-b)\tan{x\over2}\over\sqrt{ a^2-b^2}\right\}.
\end{equation}
One finds that the divergent contribution to the effective Lagrangian is given by
\begin{eqnarray}
&{}&
{\sin^2\chi\sin^2\theta\over24\pi^2r^4}\sum_{m=1}^\infty\left(m\beta\over
r\right)^{-2}\left\{{g(\beta)ma\sinh(ma)\over h(\beta)\sigma_m(x)^2}+
{X(\beta)ma\sinh(ma)\over\left(\cosh\left(m\beta\over
r\right)-1\right)h(\beta)^2\sigma_m(x)}\right. \nonumber \\
&{}&\left. -{g(\beta)\over\sin^2\chi\sin^2\theta h(\beta)\sigma_m(x)}
+{X(\beta)\ln\sigma_m(x)\over\sin^2\chi\sin^2
\theta\left(\cosh\left(m\beta\over r\right)-1\right)h(\beta)^2}\right\}
\end{eqnarray}
where
\begin{equation}
g(\beta)=4\left(\cosh\left(m\beta\over r\right)-1\right) + 8\sin^2\chi\sin^2\theta +
2\sin^2\chi\sin^2\theta\left(\cosh\left(m\beta\over r\right)+1\right)
\end{equation}
\begin{equation}
h(\beta)=\cosh\left(m\beta\over r\right)-1 + 2\sin^2\chi\sin^2\theta
\end{equation}
\begin{eqnarray}
X(\beta)&=&(g(\beta)-6h(\beta))\left(\cosh\left(m\beta\over r\right)-1 +
\sin^2\chi\sin^2\theta\right) +
2\sin^2\chi\sin^2\theta\Biggl(g(\beta)\Biggr. \nonumber \\
&-&\left.h(\beta)\left(\cosh\left(m\beta\over r\right)+5\right)+
2\left(\cosh\left(m\beta\over
r\right)-1\right)\left(\cosh\left(m\beta\over r\right)+1\right)\right)
\end{eqnarray}

Ultimately, one is interested in the most dominant divergence in the effective
Lagrangian at the polarised hypersurfaces of the acausal analytic continuation. For
our purposes, therefore, one only wants the terms that diverge at least as strongly as
${1\over\sigma_m(x)^2}$, which is how (\ref{uvpart}) behaves. All other terms,
including the finite contributions, can be neglected in future calculations without
losing any of the essential physics. Finally, therefore, one obtains
\begin{equation}
\label{fin}
\ln Z(x)={1\over4\pi^2r^4}\sum_{m=1}^\infty\left\{{\left(m\beta\over
r\right)^{-1}\sinh\left(m\beta\over r\right)\over\sigma_m(x)^2} + {\left(m\beta\over
r\right)^{-2}g(\beta)ma\sinh(ma)\over6h(\beta)\sigma_m(x)^2}\right\}
\end{equation}
as the dominant contribution. The first point to note about the second term in this
expression is that it reinforces the first term, {\it i.e.} it also diverges to
infinity at each of the polarised hypersurfaces. In fact, both terms are equal at the
point where all the polarised hypersurfaces coincide ({\it i.e.} if $a={\beta\over r}$ 
and $\sin\chi\sin\theta=1$). However, unlike (\ref{uvpart}), the second term cannot be
derived from the asymptotic expansion of the heat kernel $H(x,x',\tau)$ near $\tau=0$.

\section{The Suppression of Acausal Effects}
\label{cvfive}

Now let us consider the Einstein universe as a fixed background on which scalar
particles can exist. In this universe, one defines energy and angular momentum by
integrating the energy--momentum tensor with the appropriate Killing vector over a
spacelike surface and these quantities are conserved in that they are the same on all
surfaces. If one now puts a certain energy in scalar particles in this universe, it
will occupy a number of states given by the entropy, and if the particles are given
angular momentum, the fluid will begin to rotate and the number of states will
decrease.

The effective action density $\ln Z(x)$ for the scalar particles is given by the 
expression
(\ref{fin}), equally valid for both analytic continuations of the Euclidean section.
Here, one can interpret $\ln Z(x)$ as the grand canonical partition function for
the hot rotating radiation. The $r\rightarrow\infty$ limit gives the partition 
function for rotating
scalar radiation in flat space, and if the angular velocity parameter $a$ is small,
one obtains the partition function for a hot rigidly rotating perfect scalar fluid
(when one integrates $\ln Z(x)$ over a cylindrical volume with radius $r'=r_B$)
\begin{equation}
\ln Z={\pi^2V\over90\beta^3}{1\over\left(1-\left(ar_B\over\beta\right)^2
\right)}\space. 
\end{equation}
The partition function satisfies 
\begin{equation}
\ln Z={\cal S} - \beta(E-\Omega J)\space,
\end{equation}
and by applying the standard thermodynamic identities, one can now calculate the 
energy of
the particles at a temperature $T=\beta^{-1}$ and also the angular momentum that is 
required to make the particles rotate with an angular velocity $\Omega=a/\beta$. In the 
Einstein universe,
\begin{eqnarray}
E(x)&=&-{\partial\ln Z(x)\over\partial\beta}=\sum_{m=1}^\infty
{2\ln Z_m(x)\over\sigma_m(x)}
\left(\partial \sigma_m(x)\over\partial\beta\right) + {1\over4\pi^2
r^4}\sum_{m=1}^\infty {1\over\beta\sigma_m(x)^2}\nonumber \\
&{}& \left\{\left(m\beta\over r\right)^{-1}\sinh\left(m\beta\over r\right)
-\cosh\left(m\beta\over r\right) +{\sin^2\chi\sin^2\theta\over6}\left( m\beta\over
r\right)^{-2}{ma\sinh(ma)\over h(\beta)^2}\right.\nonumber \\
&{}&~~~~~~~~~~\Biggl.\Bigl(2g(\beta)h(\beta) -\beta
g'(\beta)h(\beta) + \beta g(\beta)h'(\beta)\Bigr)\Biggr\}
\end{eqnarray}
\begin{eqnarray}
\label{ang}
J(x)&=&{\partial\ln Z(x)\over\partial a}=-\sum_{m=1}^\infty
{2\ln Z_m(x)\over\sigma_m(x)}\left(\partial
\sigma_m(x)\over\partial a\right) + {1\over4\pi^2 r^4}\sum_{m=1}^\infty {1\over
a\sigma_m(x)^2}\nonumber \\
&{}&\left\{ {\sin^2\chi\sin^2\theta\over6}\left(m\beta\over
r\right)^{-2}{g(\beta)\over h(\beta)}\Bigl(ma\sinh(ma) + (ma)^2\cosh(ma)\Bigr)\right\}
\end{eqnarray}
where $\ln Z(x)=\sum_{m=1}^\infty \ln Z_m(x)$. These expressions diverge to infinity
at the critical angular velocity $\Omega=1/r$ (if $\sin\chi\sin\theta=1$). Physically, 
this means that one
would have to put an infinite amount of angular momentum into the system if one wanted
the particles at the boundary to move at the speed of light. If one now calculates the
entropy of the particles, then one obtains
\begin{eqnarray}
{\cal S}(x)&=&\sum_{m=1}^\infty{2\ln Z_m(x)\over\sigma_m(x)}\left[\left(m\beta\over 
r\right)\sinh\left(
m\beta\over r\right) -\sin^2\chi\sin^2\theta ma\sinh(ma)\right]
+{1\over4\pi^2 r^4}\sum_{m=1}^\infty{1\over\sigma_m(x)^2}\nonumber \\
&{}&\left\{ 2\left(m\beta
\over r\right)^{-1}\sinh\left(m\beta\over r\right) - \cosh\left(m\beta\over r\right)
-{\sin^2\chi\sin^2\theta\over6}\left(m\beta\over r\right)^{-2}{g(\beta)\over
h(\beta)}(ma)^2\cosh(ma)\right.\nonumber \\
&{}&\left.+{\sin^2\chi\sin^2\theta\over6}\left(m\beta\over r\right)^{-2}{ma\sinh(ma)
\over h(\beta)^2}\Bigl(2g(\beta)h(\beta)-\beta g'(\beta)h(\beta) + \beta
g(\beta)h'(\beta)\Bigr)\right\}
\end{eqnarray}
We want to consider what happens to the number of states with a given energy as
the angular momentum is increased. However, as one injects more and more angular
momentum into the system so that the angular velocity approaches the critical
value $\Omega=1/r$, the energy of the particles also diverges to infinity. This
means that in order to keep the energy fixed, one must decrease the fluid
temperature by an appropriate amount as the energy increases. In particular, as
the energy diverges to infinity, the temperature must be scaled to zero. Therefore, as
$\Omega$ approaches its critical value and as the parameter $\beta$ tends to
infinity, one can see that the entropy diverges to minus infinity. This means that as 
one gets nearer
and nearer to making the particles travel faster than the speed of light, their number
of states decreases to zero. 

Of course, the identified Euclidean Einstein universe can be analytically continued in a
different way to obtain $3dS\times S^1$, the product of three dimensional de Sitter
space and the $S^1$ with length $\beta$. In this case, however, the conserved 
quantities are no longer
energy and angular momentum, but rather linear momentum and boost momentum. Once
again, therefore, one can consider $3dS\times S^1$ as a fixed background containing
scalar field particles with a fixed amount of linear momentum, occupying a certain
number of states. One can give these particles boost momentum and the amount that is
needed to boost the particles to a certain value of $a$ is again determined by
the formula (\ref{ang}). The number of states available for the particles is given by
the entropy, which decreases as the particles are boosted to higher and higher values.
As was discussed in section\ \ref{cvtwo}, CTCs appear in this spacetime at the
critical value $a=\beta/r$, but the amount of boost momentum that is needed to obtain
this critical value is actually infinite, and the corresponding number of states
available to the system falls to zero as ${\cal S}$ diverges to minus infinity. In
this example, therefore, one can see that there are insuperable obstacles to 
introducing CTCs and no available quantum states for causality violation. 

Finally, let us consider the creation of $3dS\times S^1$ from nothing, a process that
can be described by constructing a no--boundary wave function.
Specifically, we want the amplitude $\Psi$ to propagate from nothing to a boosted scalar
field configuration on the three--surface with topology $S^1\times S^2$ at constant
$\phi$. One must have a fixed linear momentum around the $S^1$, characterised by the
parameter $\beta$, while the amount of boost is determined by the parameter $a$. The
wave function will be given by a Euclidean path integral of the usual form, but once
again care must be taken. The amplitude is
described by cutting the original solution in half, but it would be a mistake to
simply use the amplitude $\ln\Psi=\ln Z/2$. One can see that this would be tantamount
to employing a grand canonical description, giving the amplitude as a function of the 
fixed `potentials' $a$ and $\beta$ but as we have stressed, the correct amplitude 
should be given as a function of the conserved `charges' appropriate for a
microcanonical description. The correct amplitude $\Psi_m$ is defined as the Legendre
transform
\begin{equation}
2\ln\Psi_m=\ln Z -\beta{\partial\ln Z\over\partial\beta} - a{\partial\ln
Z\over\partial a}.
\end{equation}
The microcanonical partition function,
or density of states, is given by the quantity $\Psi_m^2$, which implies that the
entropy is just $2\ln\Psi_m$. In this example, therefore, one can see that the 
amplitude to propagate from nothing to a boosted
scalar field configuration is nonzero if the boost is not too large. As soon as it
becomes large enough to lead to the formation of CTCs, however, the amplitude vanishes
exponentially.

\section{Discussion}
\label{cvsix}

In this paper, we have considered the behaviour of a scalar quantum field on a
background spacetime whose metric parameters can be adjusted so as to introduce CTCs.
The entropy has been shown to diverge to minus infinity at the onset of causality
violation, which can be interpreted as saying that the number of available quantum
states tends to zero. The crucial question to ask, therefore, is whether this result
holds in the general case.

The key quantity in our analysis has been the effective action density, which initially
diverges to infinity. In section\ \ref{cvfour}, it was shown that the strongest
divergence in this action has two distinct contributions. The first contribution can
be derived from the asymptotic expansion of the heat kernel, and it has been shown
that in general this contribution diverges to infinity for fields of arbitrary mass
and spin at the polarised hypersurfaces of an acausal spacetime \cite{mjc}. The only
exceptions to this rule occur if the Van--Vleck determinant is made to vanish, as in
Visser's Roman ring configuration \cite{viss}. However, the second contribution to the
action cannot be derived from a knowledge of ultra--violet behaviour and need not
depend on the Van--Vleck determinant. One would expect this term to diverge with the
same sign as the other dominant contribution, so even in the case of the Roman ring the
action should still diverge to infinity, although this remains to be explicitly shown.

Many chronology violating spacetimes, including the ones considered in this paper,
are multiply connected and in general, any multiply connected acausal spacetime should 
have a simply connected
covering space with points identified under a discrete group of isometries. The
effective action density will be given as a function of the metric parameters which 
determine
the spacetime interval separating two of the identified points and in a thermodynamic
context, one can see that these parameters are just thermodynamic intensive variables,
which were interpreted as temperature and angular velocity in the rotating fluid model 
considered here. Similar parameters will also exist if one analytically continues from
some Euclidean metric to obtain a simply connected acausal spacetime. Therefore, one 
should always be able to Legendre transform the effective action
in order to calculate the density of states, which must be defined as a function of
the thermodynamically conjugate extensive variables. The evidence presented in this
paper suggests that the resulting density of states will tend to zero
as the parameters are adjusted so as to introduce CTCs. Thus it appears that quantum
mechanics naturally forbids acausal behaviour. There are no quantum states
available for causality violating field configurations.

\end{document}